\documentclass{IEEE_lsens}
\usepackage{textcomp}
\usepackage{subcaption}
\usepackage[noadjust]{cite}
\usepackage{siunitx}
\usepackage[pdftex]{graphicx}
\graphicspath{{./images/}}
\DeclareGraphicsExtensions{.pdf,.jpeg,.png}
\DeclareUnicodeCharacter{015F}{\weirdS}
\usepackage{amsmath}
\usepackage[cmintegrals]{newtxmath}
\usepackage[utf8]{inputenc}
\usepackage{makecell}
\usepackage{graphicx}
\usepackage{array}
\usepackage{hhline}
\usepackage{booktabs}

\usepackage{booktabs,threeparttable}
\usepackage{bm}
\usepackage{algorithmic}
\usepackage{array}
\usepackage{url}
\newcommand\MYhyperrefoptions{hypertexnames=true, bookmarks=true,
bookmarksnumbered=true, pdfpagemode={UseOutlines}, plainpages=false,
pdfpagelabels=true, colorlinks=true, linkcolor={black},
citecolor={black}, urlcolor={black}}
\ifCLASSINFOpdf
\usepackage[\MYhyperrefoptions,breaklinks=true,pdftex]{hyperref}
\else
 \usepackage[\MYhyperrefoptions,dvips]{hyperref}
 \usepackage{breakurl}
\fi
\usepackage{tikz}

\newcommand*\blackcircled[1]{\tikz[baseline=(char.base)]{
            \node[scale=1.0,shape=circle,draw, fill,inner sep=3pt] (char) {};
            \node[scale=1.0] (char) {\textcolor{white}{\footnotesize{\textbf{#1}}}}}}
\begin{document} 
\hypersetup{pdftitle={Open Gimbal: A 3 Degrees of Freedom Open Source Sensing and Testing Platform for Nano and Micro UAVs},
pdfsubject={Open Gimbal},
pdfauthor={Suryansh Sharma},
pdfkeywords={Nano UAV, Batteryless Sensing, Gimbal, Gyroscopic Testbed, Wireless Sensor}}
\title{Open Gimbal: A 3 Degrees of Freedom Open Source Sensing and Testing Platform for Nano and Micro UAVs}
\author{\IEEEauthorblockN{Suryansh~Sharma\IEEEauthorrefmark{1}, 
Tristan~Dijkstra\IEEEauthorrefmark{2},
and R. Venkatesha Prasad\IEEEauthorrefmark{1}\IEEEauthorieeemembermark{1}}% <-this % stops a space
\IEEEauthorblockA{\IEEEauthorrefmark{1}Networked Systems Group, Delft University of Technology, The Netherlands\\
\IEEEauthorrefmark{2}Faculty of Aerospace Engineering,
Delft University of Technology, The Netherlands\\
\IEEEauthorieeemembermark{2}Senior Member, IEEE}%
\thanks{Corresponding author: Suryansh Sharma (e-mail: Suryansh.Sharma@tudelft.nl).}
% Associate Editor: F. Costa.\\
% Digital Object Identifier 10.1109/LSENS.2023.3307121}
}
% Manuscript received line
% \IEEELSENSmanuscriptreceived{Manuscript received 26 July 2023; accepted 3 August 2023. Date of publication 22 August 2023.
% Date of publication August 12, 2023.
% }
\IEEEtitleabstractindextext{%
\begin{abstract}
Testing the aerodynamics of micro- and nano-UAVs without actually flying is highly challenging. To address this issue, we introduce \textit{Open Gimbal}, a specially designed 3 Degrees of Freedom platform that caters to the unique requirements of micro- and nano-UAVs. This platform allows for unrestricted and free rotational motion, enabling comprehensive experimentation and evaluation of these UAVs. Our approach focuses on simplicity and accessibility. We developed an open-source, 3D printable electro-mechanical  design that has minimal size and low complexity. This design facilitates easy replication and customization, making it widely accessible to researchers and developers. Addressing the challenges of sensing flight dynamics at a small scale, we have devised an integrated wireless batteryless sensor subsystem. Our innovative solution eliminates the need for complex wiring and instead uses wireless power transfer for sensor data reception. To validate the effectiveness of open gimbal, we thoroughly evaluate and test its communication link and sensing performance using a typical nano-quadrotor. Through comprehensive testing, we verify the reliability and accuracy of open gimbal in real-world scenarios. These advancements provide valuable tools and insights for researchers and developers working with mUAVs and nUAVs, contributing to the progress of this rapidly evolving field.
\end{abstract}
}
\maketitle
%\vspace{-5pt}
\section{Introduction}
%\vspace{-5pt}
Miniaturized Unmanned Aerial Vehicles (UAVs) show great promise for a wide range of applications, including greenhouse monitoring~\cite{radoglou2020compilation}, disaster management\cite{erdelj2016uav}, crowd monitoring\cite{palossi2021fully}, inventory management\cite{longhi2017flying}, and environmental monitoring\cite{allegretti2015recharging}. These UAVs have compact dimensions, enabling them to navigate through narrow spaces, while their lightweight design ensures safe flight near humans. Typically, micro-UAVs (mUAVs) have a width of around 25\,cm and weigh up to 500\,g, while nano-UAVs (nUAVs) are approximately 10\,cm wide and weigh up to 50\,g. These UAVs are powered by motors in the range of a few Watts and are primarily operated in large indoor spaces. The popularity of small UAVs has been steadily increasing, driving ongoing research in sensor payload development, control algorithm design, and onboard sensor-based stabilization techniques\cite{preiss2017crazyswarm,niculescu2022energy,csimcsek2021novel}.
Despite their agility and maneuverability, mUAVs and nUAVs face limitations in terms of sensing and processing capabilities. Their agility also makes them prone to crashes when not piloted correctly. Since these platforms are often not built for high endurance or ruggedness, crashes can result in irreparable damage. Furthermore, developing state estimation and control algorithms for multirotor mUAVs and nUAVs operating in real-world scenarios is a lengthy, iterative, and risky process. Therefore, it is essential to have testing and evaluation capabilities in controlled and safe environments. Research and development on these UAVs necessitate a test platform that allows for controlled flight testing, tuning of PID or other control parameters, studying rotor failure response, and measuring UAV performance under different conditions.
A suitable test gimbal for mUAVs and nUAVs should fulfill several requirements: \blackcircled{1} it should enable unrestricted rotational motion (3 Degrees of Freedom (DoF)), \blackcircled{2} it should add minimal additional inertia to the UAV's flight dynamics to preserve its original characteristics, \blackcircled{3} it should be compatible with small, weakly powered motors that are sensitive to even minor changes in mass, and \blackcircled{4} it should sense the attitude and accelerations of the tested platform without interfering with its dynamics or control algorithms. Though there are some previous works, they have some shortcomings too. Thus we compare relevant works from the literature in Section~\ref{comparison}. 

In this work, we introduce open gimbal: a specially designed 3 DoF platform that caters to the unique requirements of mUAVs and nUAVs, offering unrestricted rotational freedom to enable their comprehensive experimentation and evaluation. Our contribution can be summarized as follows: \blackcircled{1} We have developed an open-source\cite{Sharma_OpenGimbal_2023}, 3D printable electro-mechanical design that emphasizes minimal size and complexity. \blackcircled{2} We have devised an integrated wireless batteryless sensor subsystem to non-intrusively measure flight dynamics at a small scale. Our innovative solution eliminates the need for complex wiring and enables wireless power transfer from a distance. \blackcircled{3} We thoroughly evaluate and test the open gimbal to validate its effectiveness, communication link, and sensing performance using a typical nano quadrotor. Through comprehensive testing, we verify the reliability and accuracy of the open gimbal in real-world scenarios.
%\vspace{-10pt}
\section{System Design}
%\vspace{-5pt}
The class of UAVs that open gimbal needs to be designed for usually have small, lightweight and often core-less DC motors equipped with sub-5\,inch propellers. For instance, an indoor nano drone Crazyflie is equipped with 16\,mm DC motors with a maximum rotation speed of 26000\,RPM and a maximum generated thrust of 15.7\,g. 
These motors were rated for 1\,A of current. This imposes additional challenges regarding how much additional inertial mass the test platform is allowed to add to these already small-scale UAVs. Further, to test algorithms that address issues like rotor failure for quadrotors, the gimbal must allow for free yaw rotation. For such system when all four rotors are functioning, the gyroscopic moment generated by their different rotational speeds is typically neglected. However, in case of complete single rotor failure, this can be significant and cause a parasitic yaw spin of the drone due to imbalanced forces. Any sensor to be placed on such a test platform must be capable of operating without wires while adding the smallest possible inertial mass. The entire test platform must be optimized for weight and have all 3 degrees of freedom to emulate the flight dynamics of the test UAV while also providing sensing information about its attitude and acceleration. 
%\vspace{-15pt}
\subsection{Sensor System Design} 
%\vspace{-5pt}
\begin{figure}[t]    
    \centering
    \includegraphics[width=0.9\linewidth]{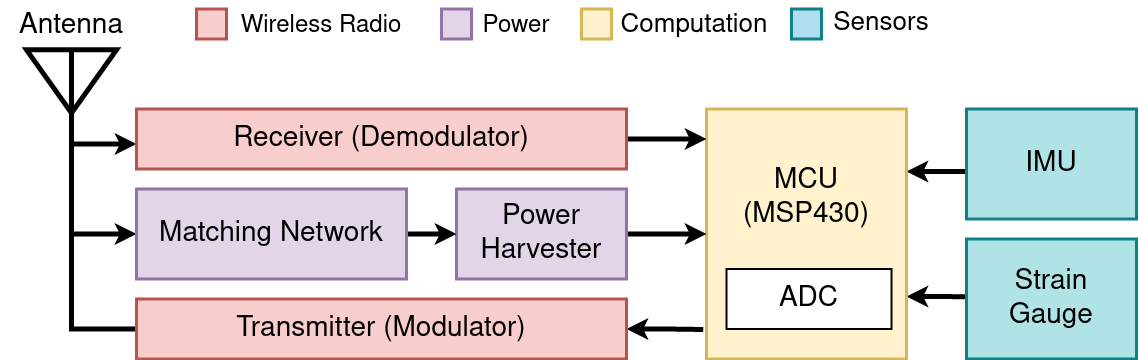}\caption{Open Gimbal: Sensing and communication Block Diagram}
    \label{fig:systemblock}
    %\vspace{-8pt}
\end{figure}
The system uses radio frequency energy harvesting and wireless backscatter techniques to power and communicates the sensed data. It is inspired by the Wireless Identification and Sensing Platform (WISP) introduced and patented by Intel\cite{sample2007design, yeager2017wisp}. The system does not require any batteries to power itself, and is environment and orientation agnostic in its operation. It includes sensors that can measure attitude and linear accelerations of the m- and nUAV under test. Fig.~\ref{fig:systemblock} shows the block diagram of the sensing system.
The system is powered by (and communicates with) an UHF Radio Frequency Identification (RFID) reader with an 8\,dBi circularly polarized reader antenna. The RFID reader acts as an interrogator and uses the EPC Class 1 Gen 2 standard to communicate with our battery-less node. A computer is connected to the RFID interrogator to send commands to the node and to receive and process the sensed data. The sensor system uses the transmitted high-frequency carrier signal to power itself, read the onboard sensors and then modulate the same carrier by either reflecting or absorbing (grounding) the signal to transmit data back through backscatter. 
To keep the weight of the sensor low, we do not add any other energy harvesting mechanisms except RF energy harvesting. For our application, the high-powered RFID transponder can be placed close to the sensor PCB which is mounted on the test gimbal. The sensor is solely powered by the RF energy harvested from the onboard antenna. The WISP platform with an identical antenna, has been shown to have a range of up to 4.5\,m\cite{yeager2017wisp}. 
After the high-power RF signal is absorbed by the antenna, it is fed into two blocks simultaneously. It is passed through a receiver which demodulates and decodes the received signal after rectification. It does so using an Onsemi NCS2200 comparator that forms an average calculator stage and has a quiescent supply current of only 10$\mu$A. At the same time, the signal is also fed to an impedance-matching network used to minimize potential reflection losses and harvest the energy. The matching network is followed by a power harvesting block that is responsible for rectifying, regulating and storing the harvested energy (in a capacitor). We use a Skyworks SMS7621 Schottky diode and a 10\,pF capacitor for rectification and a TI BQ25570 harvester IC to collect energy. This energy then triggers the ultra-low power Micro Controller Unit (MCU) and reads the sensor values. TI MSP430 was chosen since it requires significantly low power. 
\begin{figure}[t]
    \centering
    \includegraphics[width=.9\linewidth]{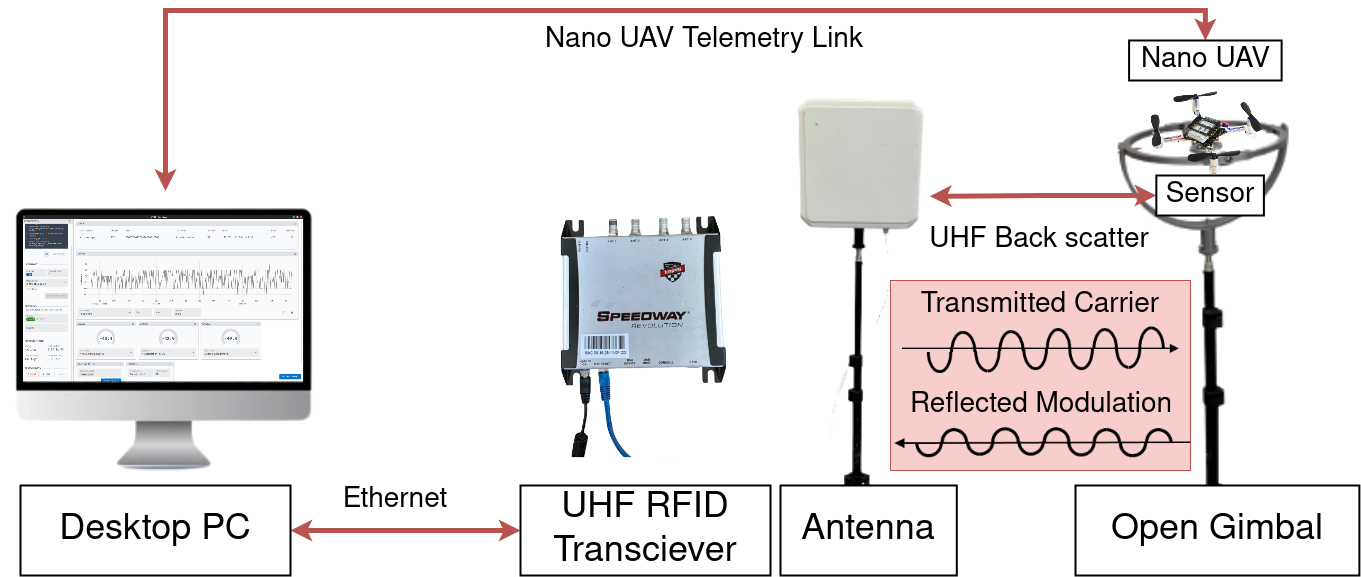}
    \caption{Open Gimbal test system}
    \label{fig:overview}
    %\vspace{-8pt}
\end{figure}
Once sufficient energy is stored in the capacitor, the MCU is activated. It uses an I2C bus to read an IMU. We use LSM303AGRTR, an ultra-low-power high-performance 3D digital linear acceleration and 3D digital magnetic sensor provide 3D attitude and acceleration values. It has a 12.6$\mu$A normal mode accelerometer current consumption. 
A sensor data packet containing the ID information (EPC identifier) based on RFID standards is created. The MCU modulates the carrier using an Analog Devices ADG902 RF Switch. The backscattered signal is then read by the RFID interrogator and further processed by the computer. The entire sensing and communication process is shown in Fig.~\ref{fig:overview}.
%\vspace{-10pt}
\subsection{Mechanical Design}
%\vspace{-5pt}
\begin{figure}[t]
    \centering
    \begin{subfigure}{0.4\linewidth}
        \centering
        \includegraphics[width=\linewidth]{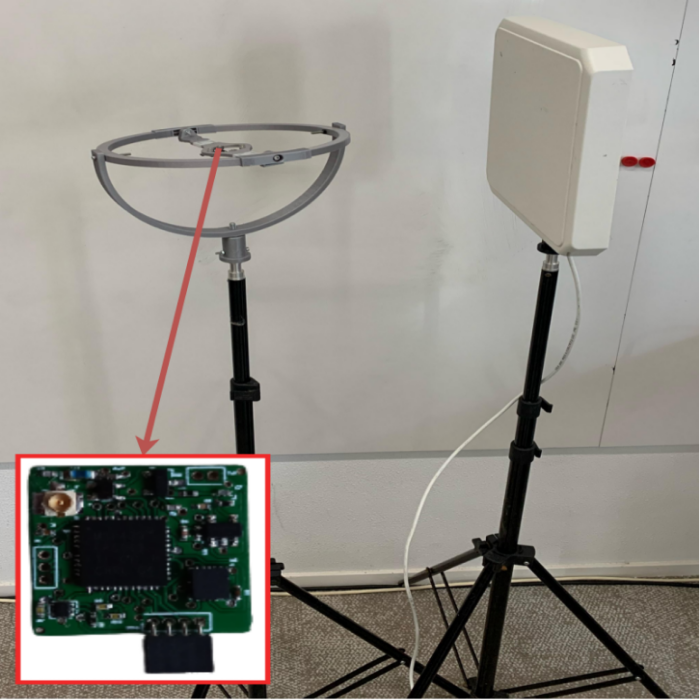}
        \caption{}
    \end{subfigure}
    \hfill 
    \begin{subfigure}{0.4\linewidth}
        \centering
        \includegraphics[width=\linewidth]{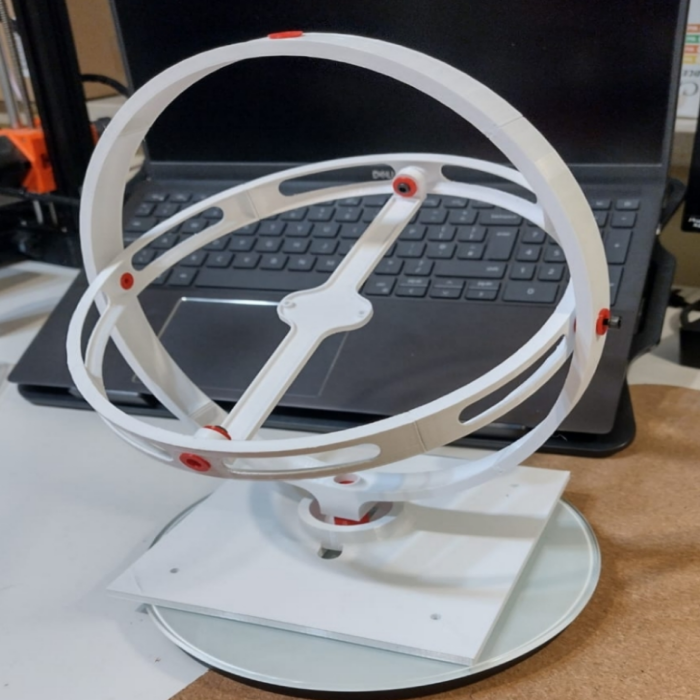}
        \caption{}
    \end{subfigure}
    \caption{Fabricated Open Gimbals for (a) Tripod and (b) Tabletop version along with the sensing system and sensor PCB.}
    \label{fig:CAD-3D-printed-sensor}
\end{figure}
The test gimbal allows full rotational degrees of freedom for the UAV under test through gyroscopic mechanism. The mechanical design consists of 3D printed parts with low-friction smooth radial bearings. Our open gimbal has 3 coupled axes: an innermost roll axis, a pitch, and an outermost yaw axis. The two types of versions, based on mounting style, are shown in Fig.~\ref{fig:CAD-3D-printed-sensor}\cite{Sharma_OpenGimbal_2023}. The sensor system is attached to the middle section of our gimbal right below the UAV mount. Mounting the entire gimbal on standard camera tripods with the screw thread of sizes 1/4-20 UNC or 3/8-16 UNC is possible. In such a case, the RFID reader for the sensor is placed laterally within a distance of 1.5\,m from the gimbal center. There is also a tabletop version where the RFID is mounted below the stand, and the entire setup is placed on a tabletop. 

\noindent\textbf{Limitations.} The added weight of the sensor is only 2\% though it can affect the dynamics, it is significantly minimal. The variations in RSSI of the packet transmission vary but for the application, it is more than sufficient. The sensor can not be used to dynamically change the orientation or control the gimbal but the idea is to get the dynamics of the drone which is possible herein. 
%\vspace{-10pt}
\section{Experimental Evaluation}
%\vspace{-5pt}
We use a popular nano UAV, Crazyflie to investigate the utility of our proposed open gimbal. Crazyflie is a nano quadrotor that flies using four 16\,mm coreless DC motors with a KV value of 14000\,RPM/V. The entire UAV measures 92\,mm diagonally and 29\,mm high. Its total weight is 27\,g~\cite{giernacki2017crazyflie}. The Crazyflie was mounted along with the sensor system on the tripod version of the gimbal. The onboard Crazyflie IMU readings were recorded using the wireless 2.4\,GHz telemetry link. The experimental maneuver consists of a 360 rotational maneuver which started with 0 initial velocity and actuated in the pitch and yaw axes. We do so by setting the maximum commanded thrust and controller setpoint for both pitch and yaw. The total maneuver time was 5\,s. The two versions of the test gimbal were 3D printed, and the completed open gimbals are shown in Fig.~\ref{fig:CAD-3D-printed-sensor}. All of the parts are 3D printed using 0\% infill setting with Polylite lightweight PLA has  a density of only 0.8\,\si{\gram\per\centi\meter\cubed}~\cite{polylite}. These weigh a total of 41.3\,g. The sensor PCB weighs 1.7\,g. The design is shared through creative commons, available at \cite{Sharma_OpenGimbal_2023}.
%\vspace{-10pt}
\subsection{Effect of orientation on communication link}
%\vspace{-5pt}
\begin{figure}[tb]
    \centering
    \begin{subfigure}{0.325\linewidth}
        \centering
        \includegraphics[width=\linewidth]{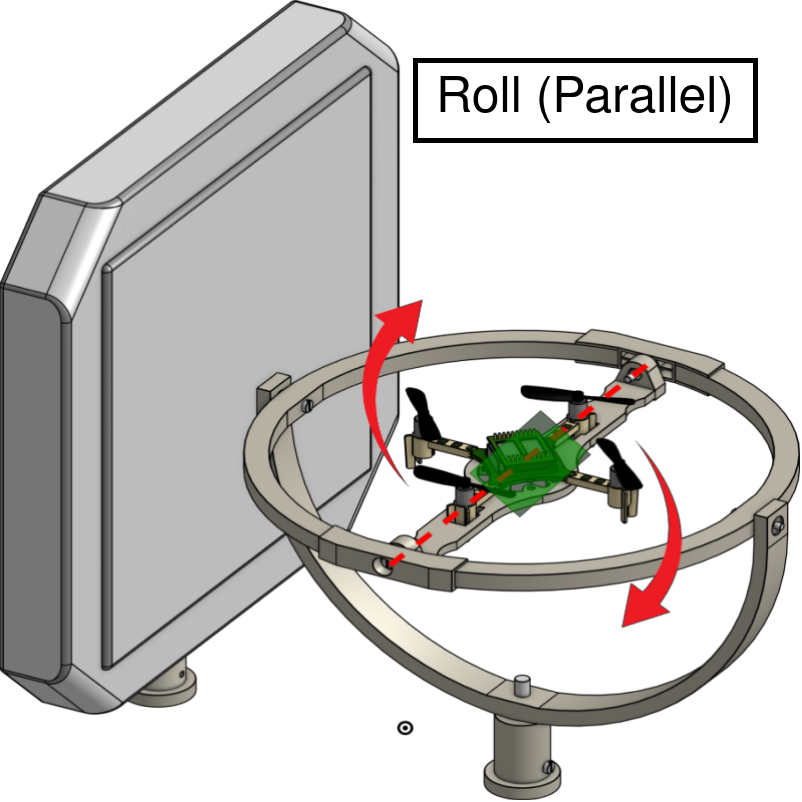}
        \caption{}
    \end{subfigure}
    \hfill 
    \begin{subfigure}{0.325\linewidth}
        \centering
        \includegraphics[width=\linewidth]{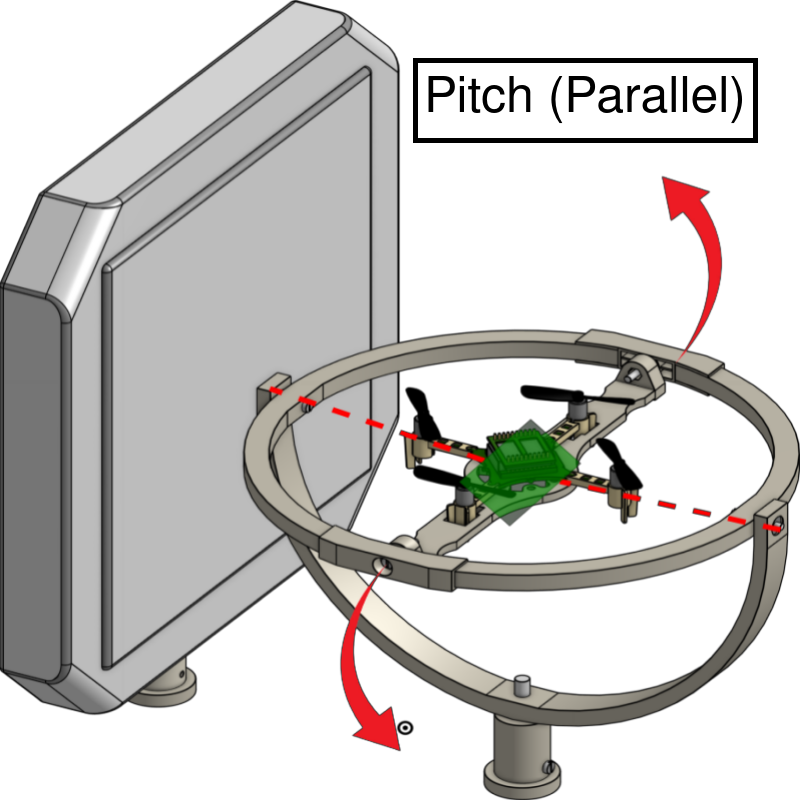}
        \caption{}
    \end{subfigure}
    \hfill 
    \begin{subfigure}{0.325\linewidth}
        \centering
        \includegraphics[width=\linewidth]{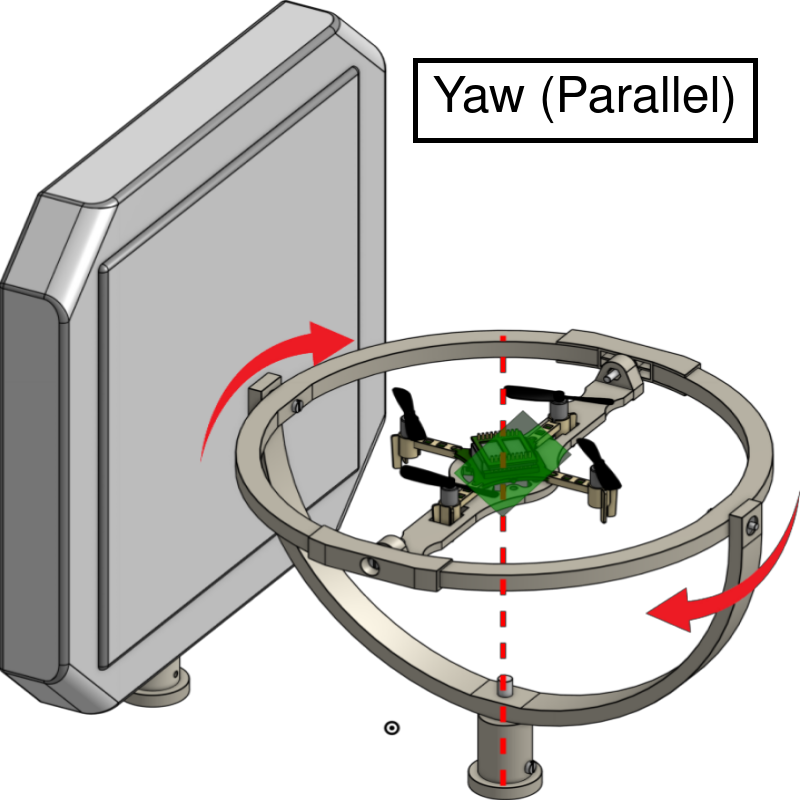}
        \caption{}
    \end{subfigure}
    \hfill 
    \begin{subfigure}{0.325\linewidth}
        \centering
        \includegraphics[width=\linewidth]{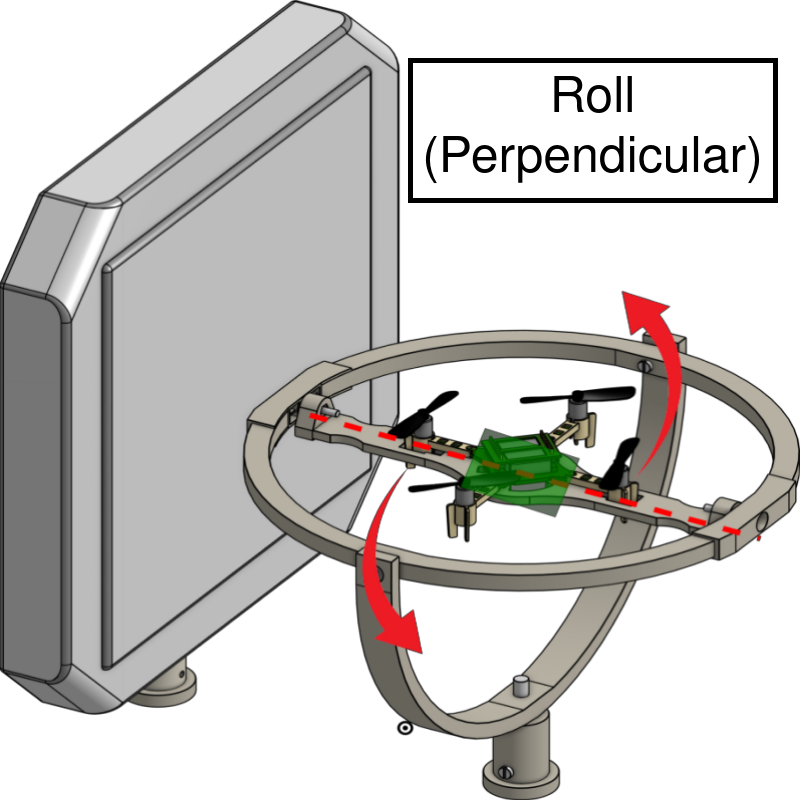}
        \caption{}
    \end{subfigure}
    \hfill 
    \begin{subfigure}{0.325\linewidth}
        \centering
        \includegraphics[width=\linewidth]{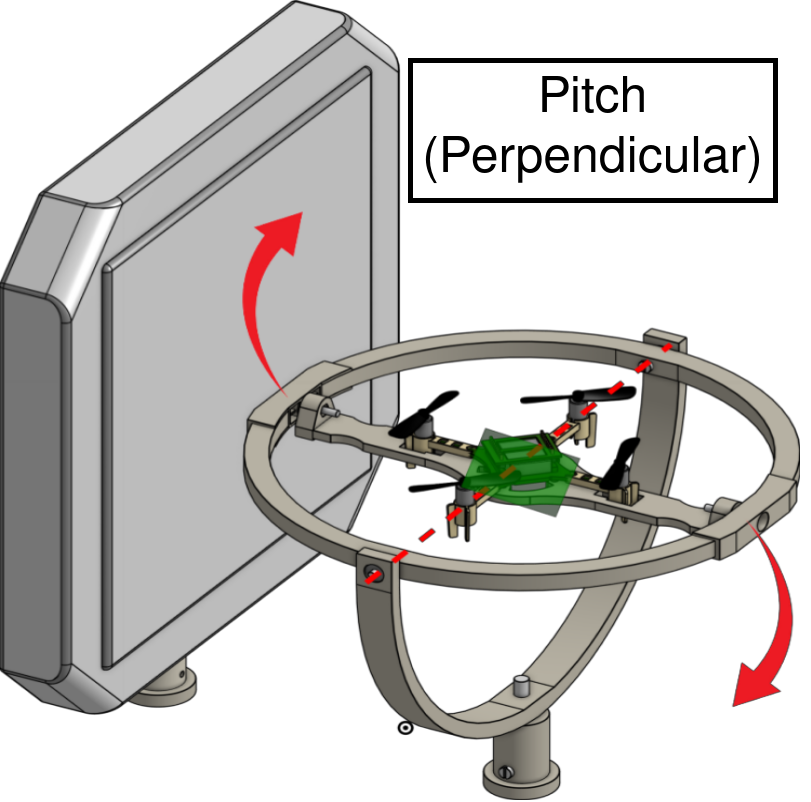}
        \caption{}
    \end{subfigure}
    \begin{subfigure}{0.325\linewidth}
        \centering
        \includegraphics[width=\linewidth]{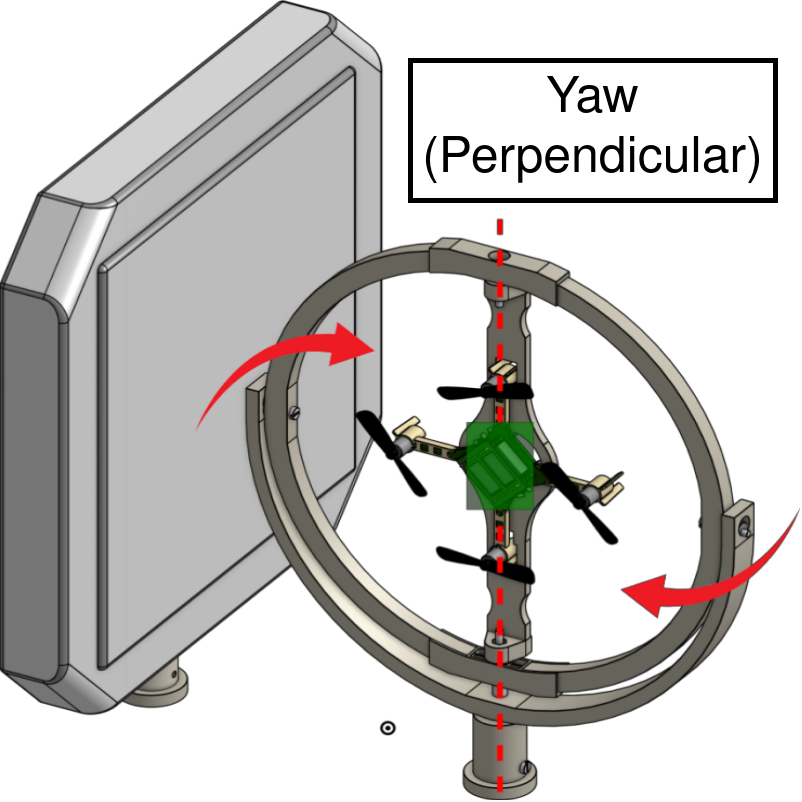}
        \caption{}
    \end{subfigure}    
\caption{Variation of sensor antenna position in roll (a,d), pitch (b,e) and yaw (c,f) axis.}
\label{fig:possible-orientations}    
%\vspace{-10pt}
\end{figure}
\begin{figure}[tb]
    \centering
    \begin{subfigure}{0.48\linewidth}
        \centering
        \includegraphics[width=\linewidth]{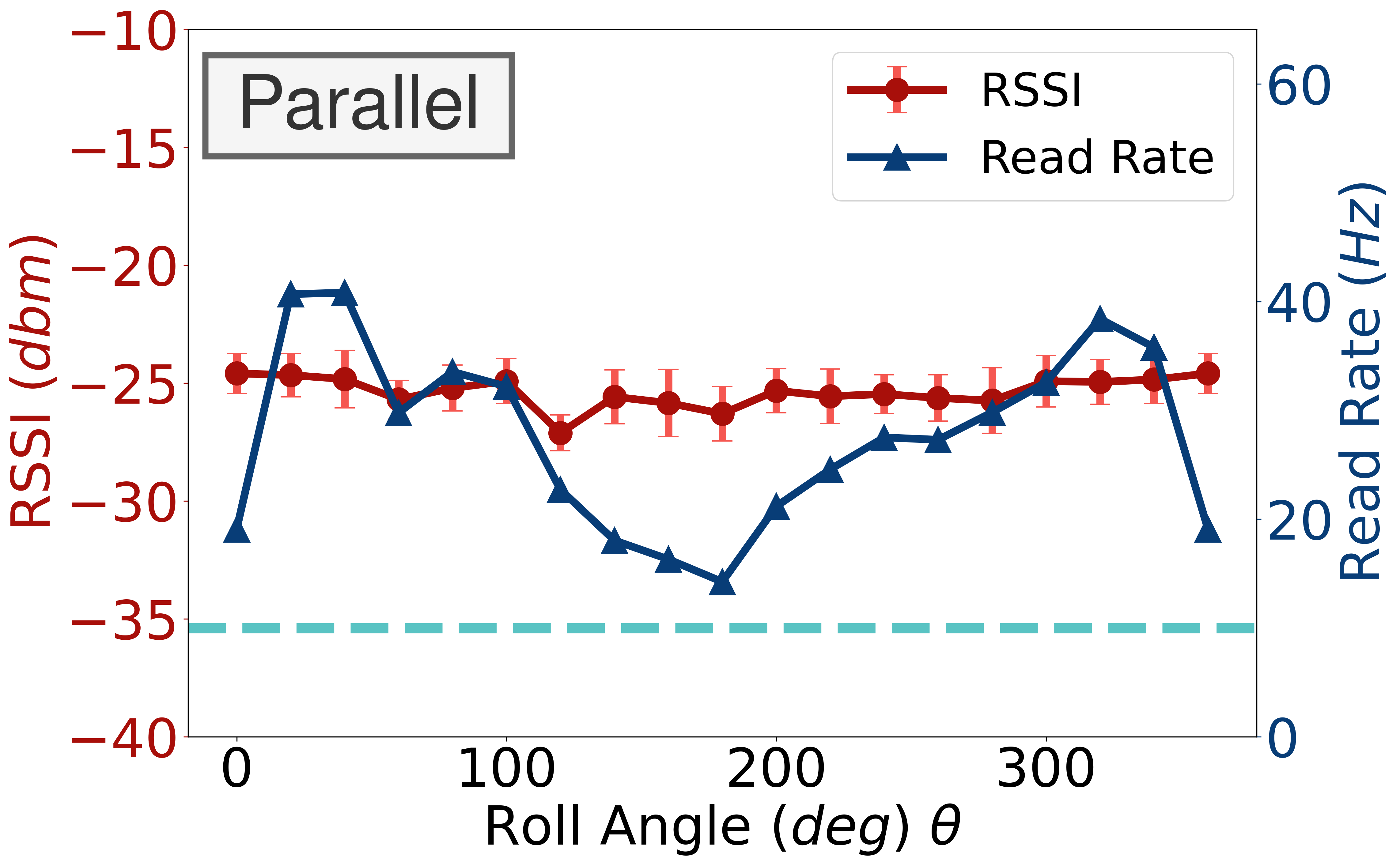}
        \caption{}
    \end{subfigure}
    \hfill 
    \begin{subfigure}{0.48\linewidth}
        \centering
        \includegraphics[width=\linewidth]{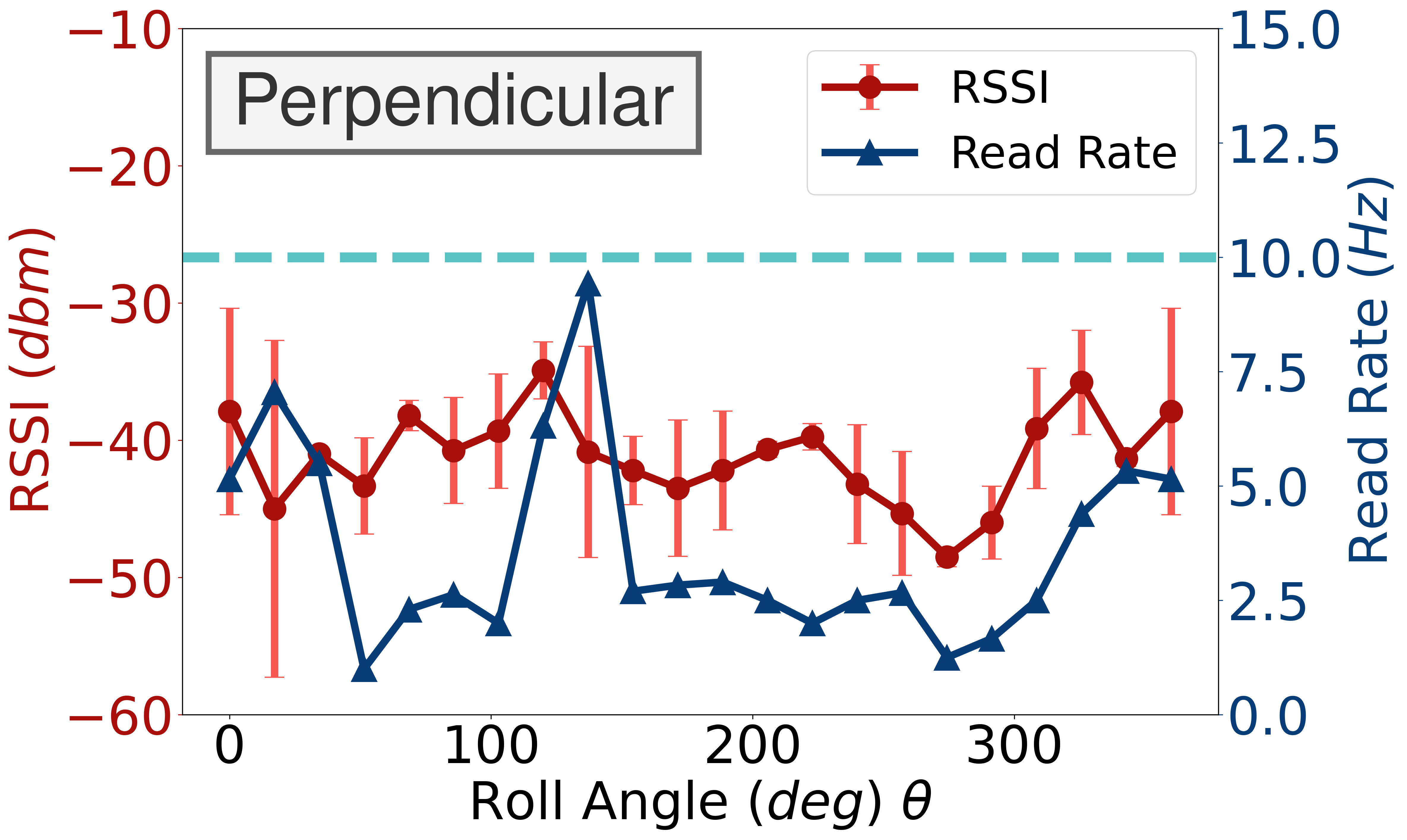}
        \caption{}
    \end{subfigure}
    \hfill 
    \begin{subfigure}{0.48\linewidth}
        \centering
        \includegraphics[width=\linewidth]{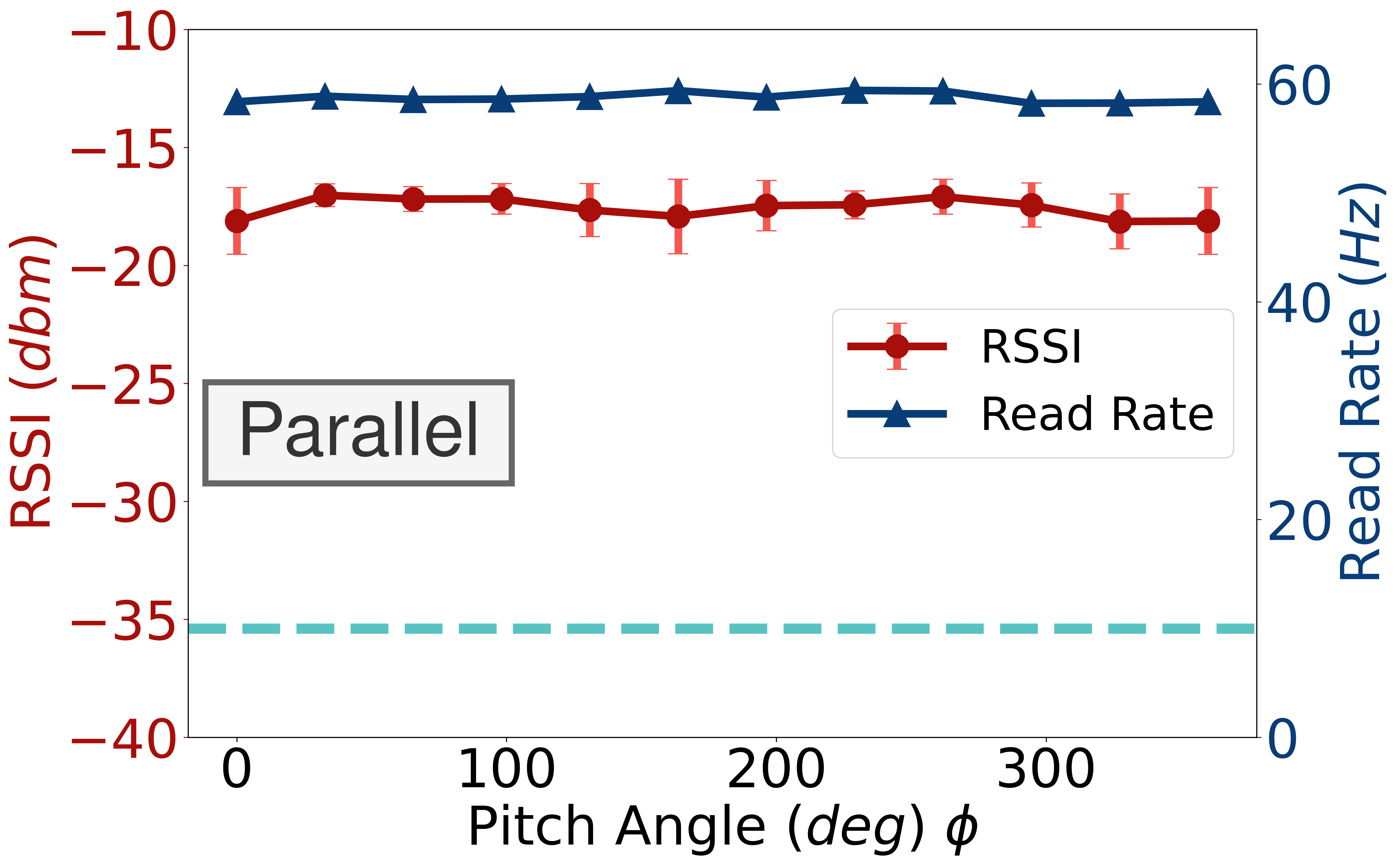}
        \caption{}
    \end{subfigure}
    \hfill 
    \begin{subfigure}{0.48\linewidth}
        \centering
        \includegraphics[width=\linewidth]{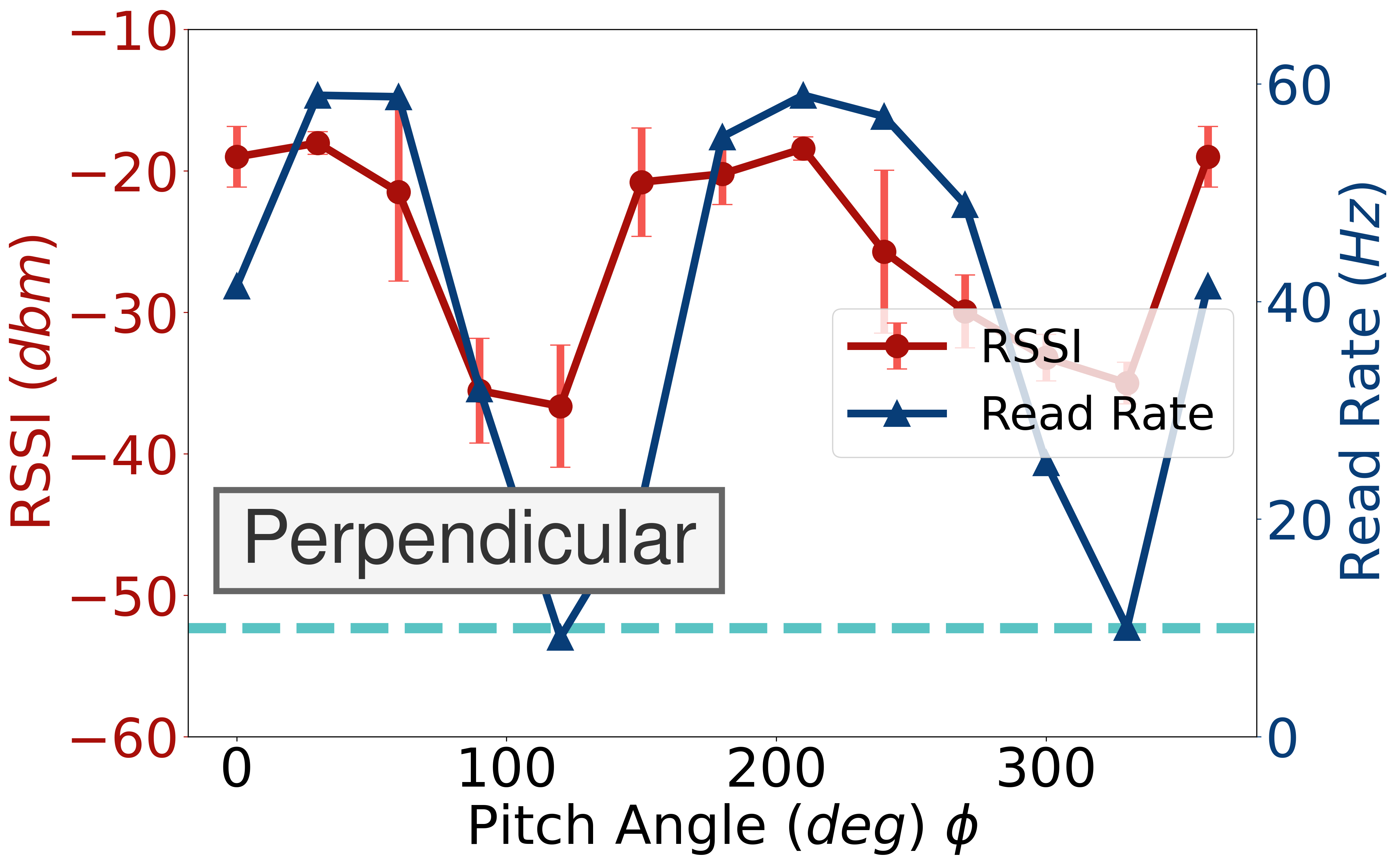}
        \caption{}
    \end{subfigure}
    \hfill 
    \begin{subfigure}{0.48\linewidth}
        \centering
        \includegraphics[width=\linewidth]{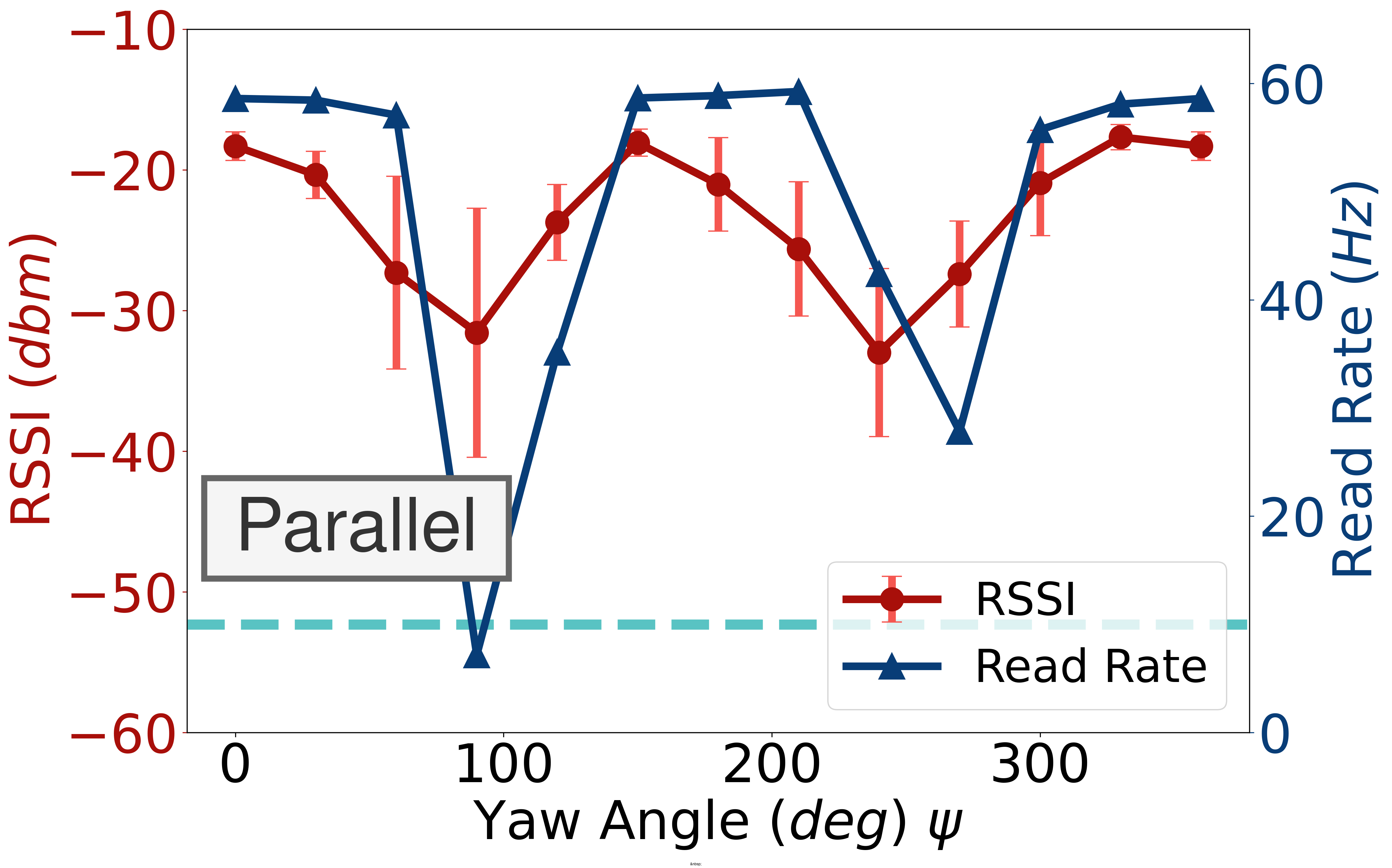}
       \caption{}
    \end{subfigure}
    \begin{subfigure}{0.48\linewidth}
        \centering
        \includegraphics[width=\linewidth]{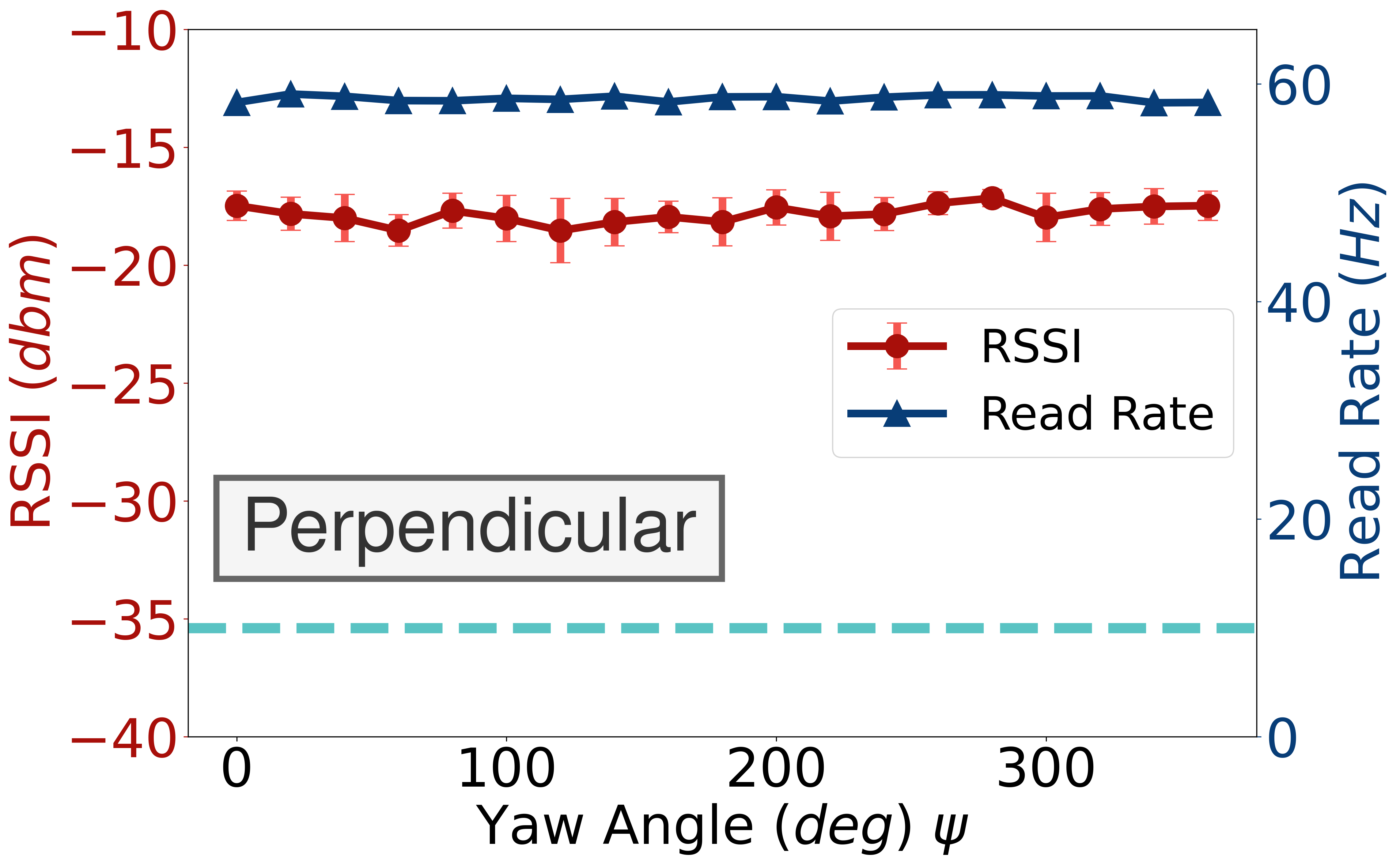}
        \caption{}
    \end{subfigure}    
\caption{RSSI of the received sensor data packet and read rate for varying yaw, pitch and roll angles for both configurations of sensor antenna position.}
\label{fig:Comm}    
\end{figure}
There are 3 independent axes in which the nUAV can freely rotate. In order to test the performance of the communication link, we considered 2 configurations of the sensor PCB with a rectangular antenna for each axis. Fig.\ref{fig:possible-orientations} shows the possible permutations. The longer side of the antenna can be perpendicular to the face of the RFID antenna (perpendicular configuration), or it can be parallel to the face(parallel configuration). This orientation will have an impact on the sensor performance on the data rate of the sensor. We characterize this in Fig.\ref{fig:Comm}. Figures (a), (c) and (e) represent the parallel configuration and (b), (d) and (f) represent the perpendicular configuration. The blue dotted line indicates the sensing rate of the wireless link from the nUAV. Communication can be established for all orientations with a much higher read rate compared to the nUAV baseline. It is interesting to note that the read rate significantly drops for the configuration perpendicular to the Roll. This can be explained by the antenna orientation being in a narrow field of view from the RFID transponder. If required, an additional RFID antenna placed perpendicularly to the first can provide a high read rate in all possible orientations of the open gimbal. 
Since the packet transfer happens on-demand using energy harvesting, we consider the RSSI to show the efficacy of open gimbal; maximum RSSI is 17.56\,dbm and minimum is -41.22\,dbm which is sufficient to receive packet transfer. Similarly, the minimum datarate is 1\,packet/s and the maximum is 59.4\,packet/s. 
%\vspace{-10pt}
\subsection{Attitude and Acceleration Sensing Performance}
%\vspace{-5pt}
\begin{figure}[tb]
    \centering
    \begin{subfigure}{0.325\linewidth}
        \centering
        \includegraphics[width=\linewidth]{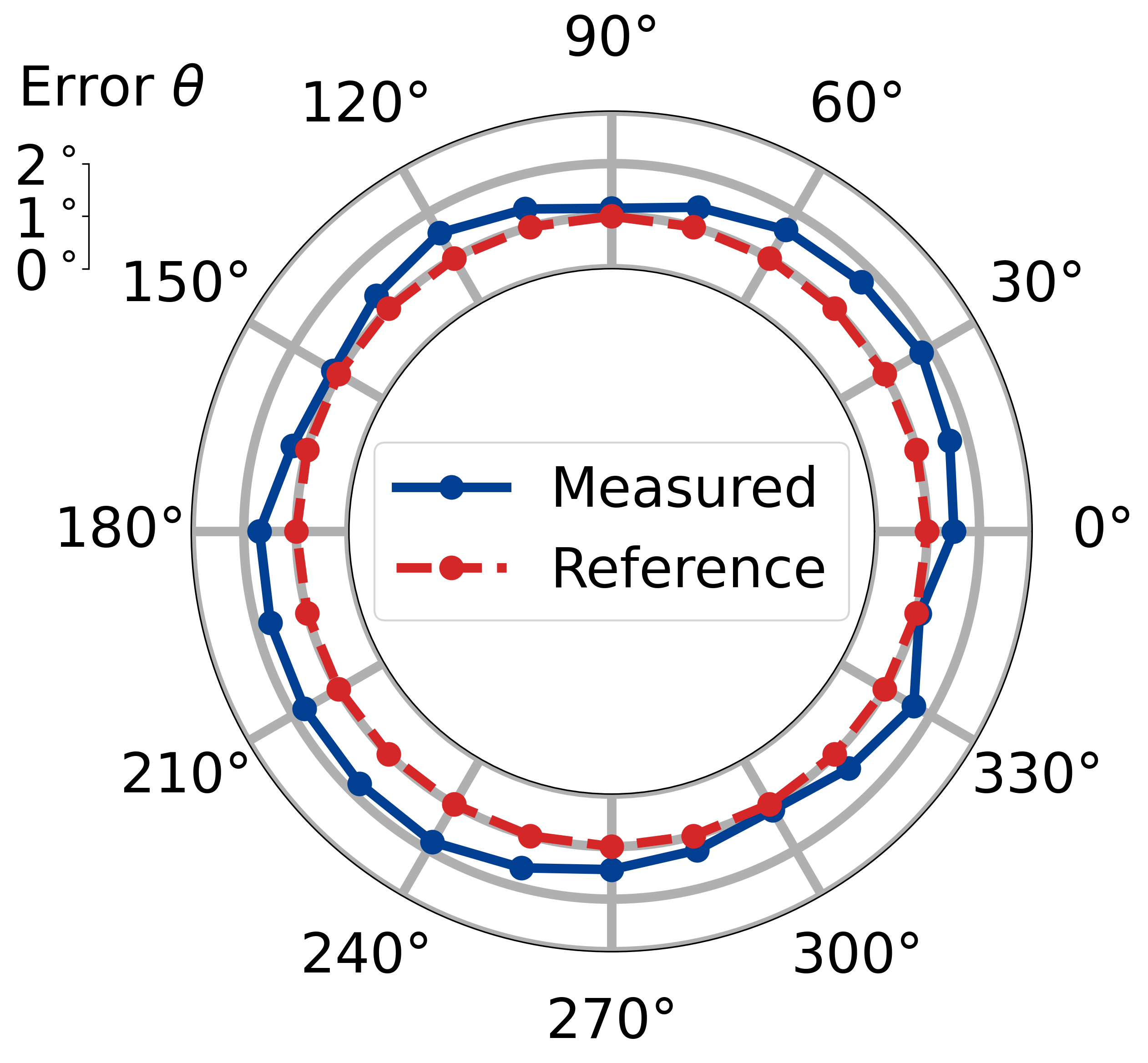}
        \caption{}
    \end{subfigure}
    \hfill 
    \centering
    \begin{subfigure}{0.325\linewidth}
        \centering
        \includegraphics[width=\linewidth]{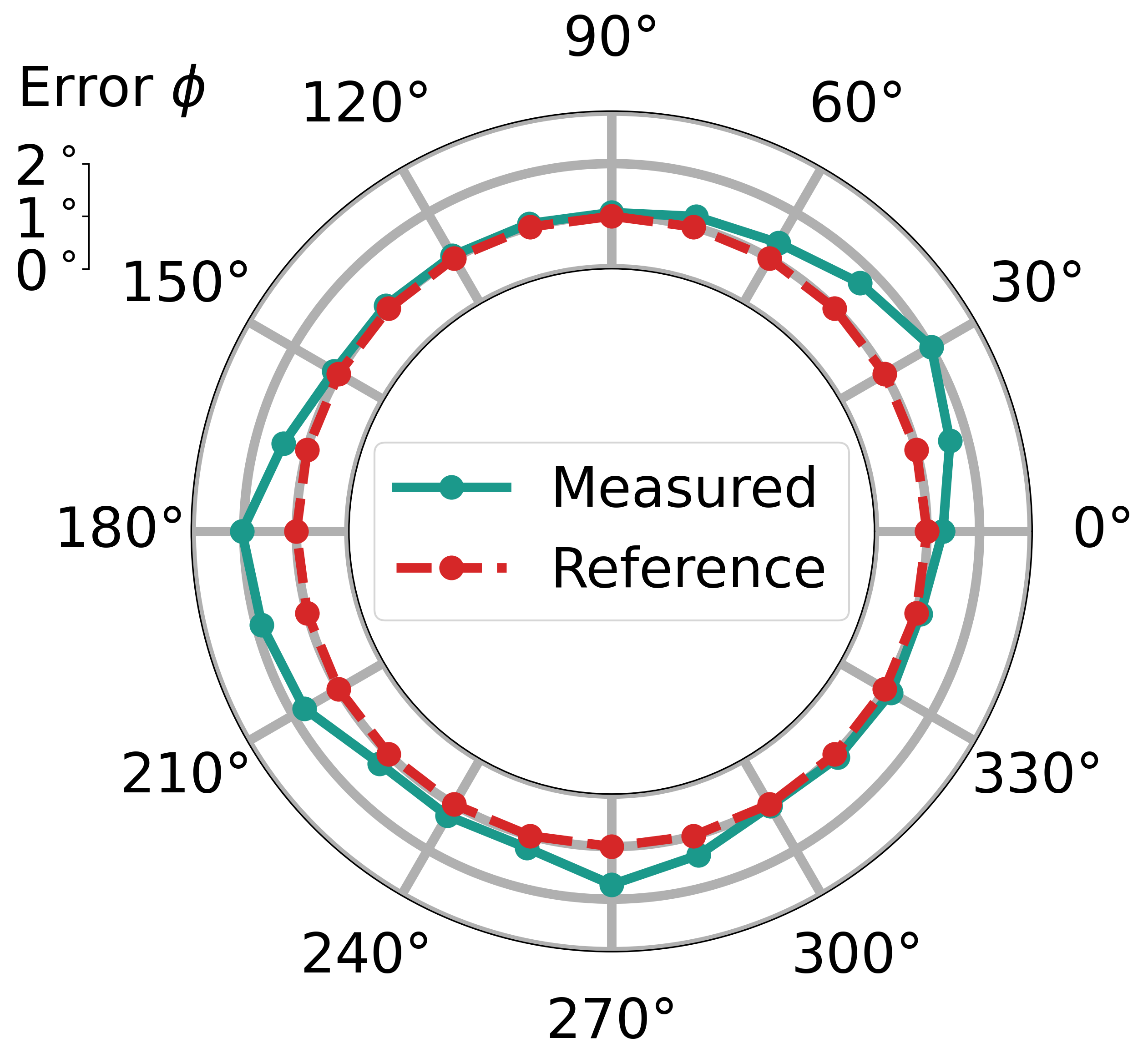}
        \caption{}
    \end{subfigure}
    \hfill 
    \centering
    \begin{subfigure}{0.325\linewidth}
        \centering
        \includegraphics[width=\linewidth]{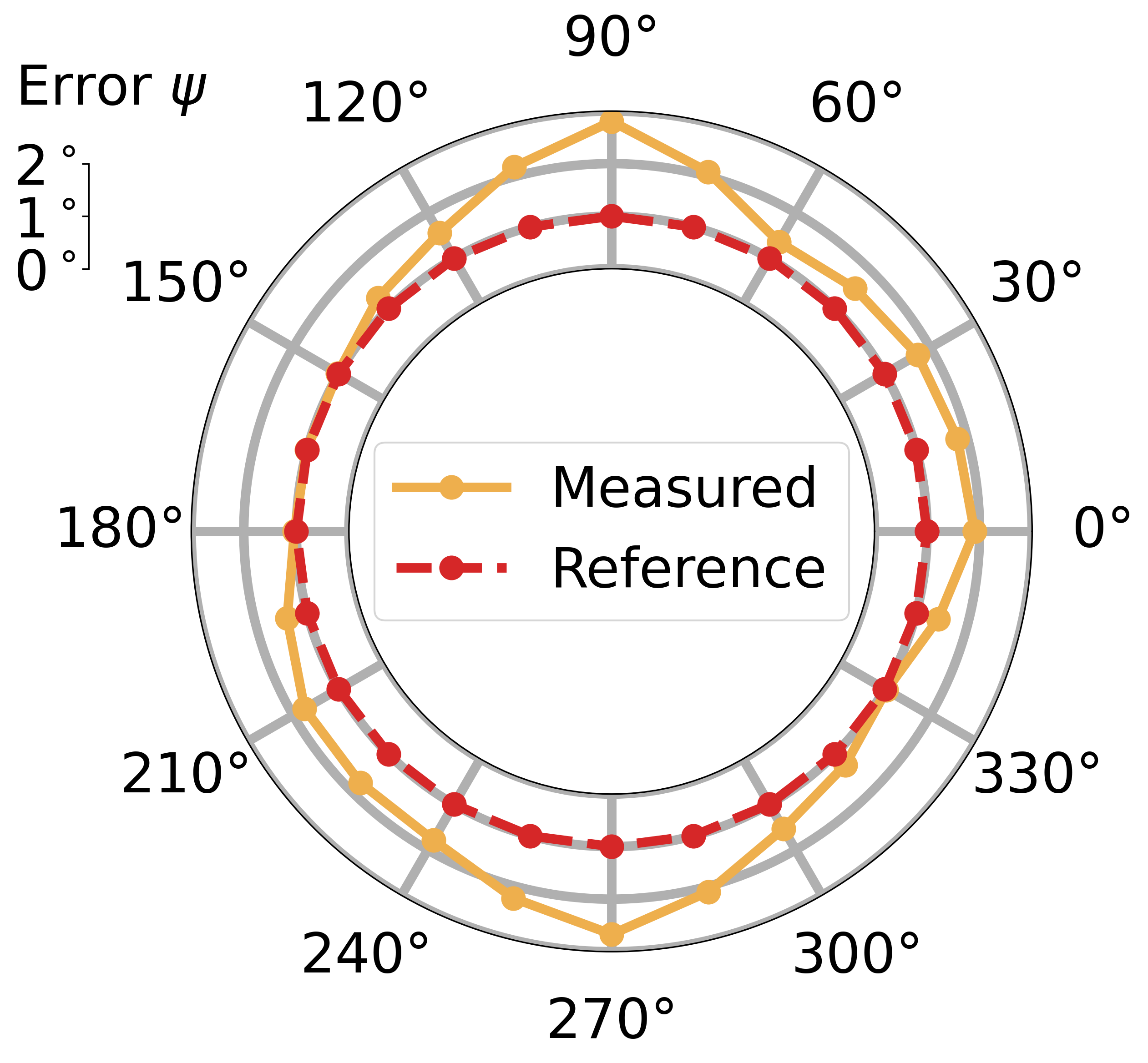}
        \caption{}
    \end{subfigure}
\caption{Attitude estimation error with a controlled reference input signal for roll, pitch, yaw ($\phi, \theta, \psi$)}
\label{fig:Attitude}
%\vspace{-12pt}
\end{figure}
\begin{figure}[tb]
    \centering
        \includegraphics[width=0.8\linewidth]{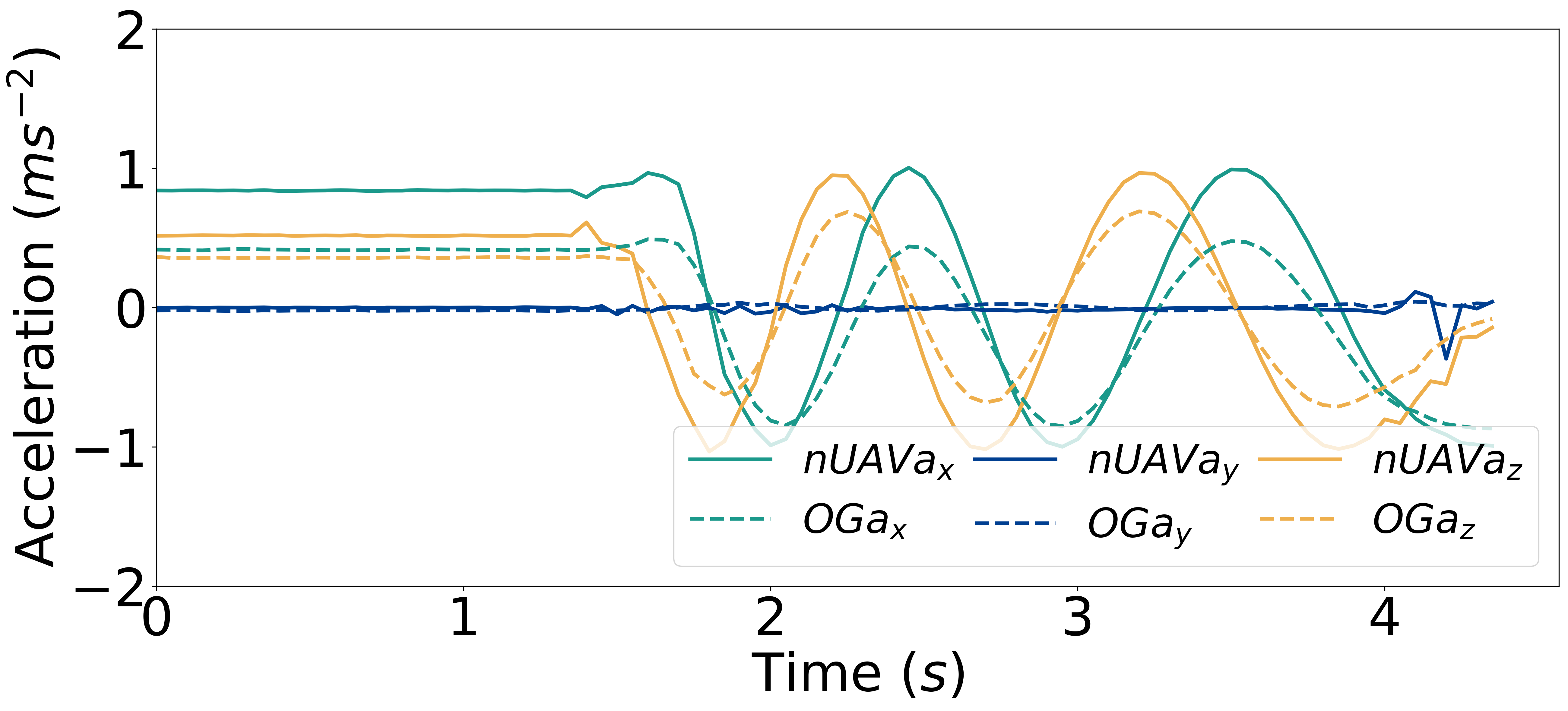}
\caption{Measured $a_x , a_y, a_z$ by the nUAV and open gimbal.}
\label{fig:Accel}
%\vspace{-10pt}
\end{figure}
The attitude of the open gimbal was changed for each independent rotational axis at a time to record the estimated attitude. This was varied into $15^\circ$ steps. The results of the attitude estimation are shown in Fig.\ref{fig:Attitude}. The absolute estimation error for all three rotational angles is within $2^\circ$. The performance of acceleration measurement of the open gimbal is shown in Fig.\ref{fig:Accel}. $360^\circ$ controlled rotations were performed, and the accelerations $a_x, a_y, a_z$ were recorded. The data was post-processed using a rolling average uniform kernel. The open gimbal data follows the same trend as reported by the nUAV with the repeatability error being 0.5\, ms$^{-2}$ and $2^\circ$ for acceleration and angle, respectively.
%\vspace{-5pt}
\subsection{Comparison of open gimbal with existing work}
\label{comparison}
%\vspace{-5pt}
Many multi-rotor testbeds have been in use since the beginning of UAV research, like the OS4 3DoF micro vertical take-off and landing (VTOL) platform ~\cite{bouabdallah2004design} or the custom quadrotor test rig built for attitude stabilization~\cite{hoffmann2010attitude}. 
DronesBench~\cite{daponte2017dronesbench} can measure thrust, power consumption, and instantaneous attitude of a light VTOL multirotor, while another custom mUAV platform uses a tri-axial gimbal with encoders mounted on every axis~\cite{bondyra2017experimental}. 
However, these platforms have mechanical limitations that restrict the complete rotational space of the UAV under test which is problematic when evaluating highly agile maneuvering like 180$\circ$ flips or rotor failure scenarios. In contrast, 3 DoF gyroscopic platform allows complete rotational experimentation~\cite{veyna2021quadcopters, santos2018experimental}. The comparison is detailed in Table 1.
\begin{table}
\centering
\caption{Comparison of Open Gimbal with literature}
\label{tab:comparison}
%\vspace{-5pt}
\resizebox{\columnwidth}{!}{%
\begin{tabular}{|>{\centering\hspace{0pt}}m{0.112\linewidth}|>{\centering\hspace{0pt}}m{0.144\linewidth}|>{\centering\hspace{0pt}}m{0.194\linewidth}|>{\centering\hspace{0pt}}m{0.206\linewidth}|>{\centering\hspace{0pt}}m{0.181\linewidth}|>{\centering\arraybackslash\hspace{0pt}}m{0.162\linewidth}|} 
\toprule
\textbf{Work}     & \textbf{360° }\par{}\textbf{Freedom} & \textbf{Sensed }\par{}\textbf{ Parameters}            & \textbf{Sensing }\par{}\textbf{ Modality}         & \textbf{Data }\par{}\textbf{ Acquisition} & \textbf{Suitability }\par{}\textbf{ n/m-UAVs}  \\ 
\hhline{|======|}
{[}14]            & No                                   & Attitude, \par{}Acceleration                          & IMU, \par{}Potentiometer                          & Wired                                     & Only \par{} VTOLs                              \\ 
\hline
{[}15]            & No                                   & Attitude, \par{}Acceleration                          & IMU, \par{}Potentiometer                          & Wired                                     & No                                             \\ 
\hline
{[}16]            & No                                   & Attitude, \par{}Thrust, \par{}Power \par{}Consumption & IMU, \par{}Load cells, \par{}Current \par{}sensor & Wired                                     & Only \par{} VTOLs                              \\ 
\hline
{[}17]            & No                                   & Attitude                                              & Tri axial \par{} encoders                         & Wireless                                  & yes                                            \\ 
\hline
{[}18]            & Yes                                  & Attitude                                              & UAV's \par{} Sensors                              & Wired                                     & No                                             \\ 
\hline
{[}19]            & Yes                                  & Only \par{} Mechanical                                & None                                              & Wired                                     & No                                             \\ 
\hline
Open~\par{}Gimbal & Yes                                  & Attitude, \par{}Acceleration                          & IMU                                               & Wireless \par{}backscatter                & Yes                                            \\
\bottomrule
\end{tabular}
}
%\vspace{-5pt}
\end{table}
%%%%%%%%%%%%%%%%%%%%%%%%%%%%%%%%%%%%%%%%%%%%%%%%%
%
%\vspace{-10pt}
\section{Conclusion}
%\vspace{-5pt}
Tiny UAVs offer significant potential for various applications due to their small size and maneuverability. However, they are prone to crashes and thus necessitate a proper testing and development platform. In this study, we introduced the design and functionality of our non-restrictive 3 DoF open gimbal platform, tailored explicitly for tiny UAVs. We demonstrated using our sensor subsystem design to measure various flight and attitude parameters. Our evaluation confirms high sensing accuracy, with attitude measurements exhibiting an error margin within $2^\circ$ and linear accelerations within 0.5$ms^{-2}$. Additionally, we assess the sensor subsystem's performance under diverse orientations and configurations of the open gimbal, achieving a consistent sensing rate of $10Hz$. Furthermore, the sensor subsystem can be expanded to accommodate additional sensors as per specific UAV testing requirements. 
The open gimbal, with its easily reproducible and customizable nature, will facilitate accelerated research and development of mUAV and nUAV platforms.
% \bibliographystyle{IEEEtran}
% \bibliography{reference.bib}
% Before submitting to IEEE Sensors Letters, manually copy in the
% resultant .bbl file contents in place of the \bibliographystyle and
% \bibliography lines here:

\section*{Acknowledgment}
% addcontentsline needed when using bookmark hyperlinking, etc.
\addcontentsline{toc}{section}{Acknowledgment}
% enable scriptsize
\scriptsize
This work has been partially supported by the H2020 ECSEL EU Project “Intelligent Secure Trustable Things” (InSecTT), funded by the ECSEL Joint Undertaking (JU) under grant agreement No 876038. The document reflects only the author’s view, and the Commission is not responsible for any use that may be made of the information it contains.

% put at least one blank line to end the scriptsize paragraph and
% then revert back to normalsize.

\end{document}